\documentclass[JHEP,12pt]{article}

\usepackage{graphics,color,soul}
\usepackage{graphicx}
\usepackage{amssymb}
\usepackage[export]{adjustbox}
\usepackage{jheppub}
\usepackage{lmodern,mathtools}
\usepackage{caption}
\usepackage{subcaption}
\usepackage{booktabs}
\usepackage[english]{babel}
\usepackage{amsmath,amssymb,amsbsy,amstext, amsthm, simplewick}
\usepackage{hyperref}
\usepackage{tikz}
\usepackage{physics}
\usetikzlibrary{decorations.markings,decorations.pathmorphing, patterns,shapes}
\usepackage{xcolor}

\usepackage{caption}

\usepackage{amsfonts}
\usepackage{amssymb}
\usepackage{upgreek}
\usepackage{simplewick}
 \usepackage{exscale,relsize}
\usepackage{mathtools}
\usepackage{comment}

\usepackage[margin=1cm,labelfont={sf,bf,scriptsize},textfont={sf,scriptsize}]{caption}

\def\be#1\ee{\begin{align}#1\end{align}}

\begin{document}

\title{Black holes from chaos
}
\affiliation{Center for the Fundamental Laws of Nature, Harvard University, Cambridge, MA, USA}

\author{Matthew Dodelson}

\abstract{We study the emergence of black hole geometry from chaotic systems at finite temperature. The essential input is the universal operator growth hypothesis, which dictates the asymptotic behavior of the Lanczos coefficients. Under this assumption, we map the chaotic dynamics to a discrete analog of the scattering problem on a black hole background. We give a simple prescription for computing the Green's functions, and explore some of the resulting analytic properties. In particular, assuming that the Lanczos coefficients are sufficiently smooth, we present evidence that the spectral density is a meromorphic function of frequency with no zeroes. Our formalism provides a framework for accurately computing the late time behavior of Green's functions in chaotic systems, and we work out several instructive examples.
}
\maketitle

\section{Introduction}
\indent The long time behavior of thermal correlation functions in chaotic theories remains an open problem in all but a few isolated examples. The simplest possibility is exponential decay,
\begin{align}\label{qnmexpansion}
\langle \mathcal{O}(t)\mathcal{O}(0)\rangle_{\beta}\sim \sum_{n}d_n e^{-i\omega_n t},\hspace{10 mm}t\to\infty,
\end{align}
where the $\omega_n$'s are a discrete set of resonances with $\text{Im }\omega_n<0$. In \cite{dodelson2024ringdownsykmodel} these resonances were studied in the context of the Sachdev-Ye-Kitaev model \cite{Maldacena:2016hyu,kitaev,Sachdev_1993,Polchinski:2016xgd}, where the correlator can be accurately computed for short time scales to thousands of digits. One then fits to exponentials to extract the resonances $\omega_n$. This level of precision is a luxury that is unrealistic in generic models, so it is beneficial to search for a method that can reproduce the same results with less input.  \\
\indent In particular, one wants to be able to accurately predict the lowest resonance from a small set of short time data, such as the first few derivatives of the correlator at time $t=0$. A similar question can be asked in the context of classical chaos, where correlation functions also exhibit resonances \cite{ruelle1990chaotic,Pollicott1985}. In that case one can obtain accurate estimates of the resonances by truncating the periodic orbit expansion to the first few shortest orbits \cite{ChaosBook,Gaspard_1998,PhysRevLett.61.2729}. Here chaos is actually an advantage rather than an obstacle, since it leads to quick convergence of the sum over orbits. Therefore there is some hope that an analogous method exists for thermal correlators in chaotic theories.\\
\indent In this paper we try to construct such a method. Motivated by the resemblance between (\ref{qnmexpansion}) and the quasinormal mode expansion in a black hole background \cite{Berti:2009kk}, we rewrite the computation of the Green's function as a scattering problem on a black hole. Roughly speaking, the radial coordinate of the black hole maps to the complexity of a given operator. The scattering potential can then be constructed out of the Lanczos coefficients, which are related to the derivatives of the correlation function at time $t=0$. Finally, the resonances are computed by solving the scattering problem. In particular, we give an example where the first four time derivatives of the correlator yield the first resonance with percent accuracy.
\\\indent This paper is organized as follows. In Section \ref{emergentsec} we introduce the Lanczos construction and its relation to black holes. In Section \ref{greenssec} we give a prescription for computing real time Green's functions. Sections \ref{freesec} and \ref{spectralsec} develop a perturbative expansion aimed at studying the analytic structure of the Green's functions. Finally, we present two examples in Section \ref{examplesec}.
\section{Emergent geometry on the Krylov chain}\label{emergentsec}
In this section we first review the Krylov chain and its relation to chaos. We then present an interpretation of the chain as the radial direction of a black hole.
\subsection{The universal operator growth hypothesis}
We consider the Hilbert space of operators $|\mathcal{O})$, equipped with the finite temperature inner product 
\begin{align}\label{innerprod}
(A|B)=\frac{\text{Tr}(e^{-\beta H}\{A^\dagger ,B\})}{2\text{Tr}(e^{-\beta H})},\hspace{10 mm}\beta=\frac{1}{T}.
\end{align}
The anticommutator Green's function of a Hermitian operator $\mathcal{O}$ is given by
\begin{align}
G_+(t)&=\frac{\text{Tr}\left(e^{-\beta H}\{\mathcal{O}(t),\mathcal{O}(0)\}\right)}{2\text{Tr}(e^{-\beta H})}=(\mathcal{O}|e^{-i\mathcal{L}t}|\mathcal{O}),
\end{align}
where the Liouvillian operator is defined by $\mathcal{L}=[H,\cdot]$.\\
\indent The Lanczos algorithm \cite{Viswanath1994TheRM,Nandy:2024htc} constructs an infinite set of operators $|\mathcal{O}_n)$ with increasing complexity. They are generated by acting with powers of the Liouvillian on the initial state $|\mathcal{O}_0)= |\mathcal{O})$ and then choosing an orthonormal basis. In this way one obtains the recursion relation 
\begin{align}
|\mathcal{O}_n)=b_n^{-1}(\mathcal{L}|\mathcal{O}_{n-1})-b_{n-1}|\mathcal{O}_{n-2})),
\end{align}
where the Lanczos coefficients $b_n$ are chosen so that $(\mathcal{O}_n|\mathcal{O}_n)=1$.\\
\indent In general not much is known about the $b_n$'s. However, in a chaotic theory it is believed that they always behave linearly at large $n$ in the thermodynamic limit \cite{Parker:2018yvk}, 
\begin{align}\label{uogh}
b_n\sim \frac{\pi}{\beta_0}n,\hspace{10 mm}n\to \infty.
\end{align}
This is known as the universal operator growth hypothesis. Assuming that the $b_n$'s are sufficiently smooth in $n$, an equivalent statement is that the spectral density exponentially decays at large frequency \cite{Avdoshkin:2019trj},
\begin{align}\label{expdecayg12}
G_+(\omega)\sim \exp\left(-\frac{\beta_0|\omega|}{2}\right),\hspace{10 mm}\omega\to \pm \infty.
\end{align}
The parameter $\beta_0$ therefore has an interpretation as an effective inverse temperature which suppresses high energy contributions to the spectral density.
\subsection{The continuum theory}
We now consider the wavefunctions
\begin{align}
\phi(n,t)=i^{-n}(\mathcal{O}_n|\mathcal{O}(t)).
\end{align}
The equations of motion in frequency space are
\begin{align}\label{eom}
i\omega\phi_\omega(n)&=b_{n+1}\phi_\omega(n+1)-b_n\phi_\omega(n-1).
\end{align}
Multiplying both sides by $\omega$ \cite{Erdmenger:2023wjg}, 
\begin{align}
\omega^2\phi_\omega(n)&=-b_{n+1}b_{n+2}\phi_\omega(n+2)+(b_{n+1}^2+b_n^2)\phi_\omega(n)-b_nb_{n-1}\phi_\omega(n-2).
\end{align}
This is a discrete wave equation in one spatial dimension labeled by $n$.\\
\indent In order to take the continuum limit, let us suppose that $\phi_\omega(n)$ and $b_n$ are sufficiently smooth at large $n$ \cite{Erdmenger:2023wjg,Muck:2022xfc,Bhattacharjee:2022vlt}. The equation of motion to lowest order in the derivative expansion then becomes 
\begin{align}\label{eomcont}
(\partial_z^2+\omega^2)\psi_\omega(z)=0.
\end{align}
Here we have defined the coordinate $z$ and rescaled field $\psi_\omega$ by
\begin{align}
dz=\frac{dn}{2 b_n},\hspace{10 mm}\psi_\omega(n)=\sqrt{b_n}\phi_\omega(n).
\end{align}
\indent Until now, we have only assumed that $b_n$ and $\phi_\omega(n)$ are smooth functions of $n$. We now add the additional input that the theory is chaotic, so that the $b_n$'s exhibit the linear growth (\ref{uogh}) at large $n$. Under this assumption, the coordinate $z$ has the asymptotic form 
\begin{align}\label{tortoisechain}
z\sim \frac{\beta_0}{2\pi}\log n,\hspace{10 mm}n\to \infty.
\end{align}
This is reminiscent of the relation between the tortoise coordinate and the radial coordinate in a black hole with temperature $\beta_0$. Assuming spherical symmetry for simplicity, the metric of such a black hole takes the form
\begin{align}
ds^2=-f(r)\, dt^2+\frac{dr^2}{f(r)}+r^2d\Omega^2,\hspace{10 mm}f'(r_s)=\frac{4\pi}{\beta_0}.
\end{align}
Here $r_s$ is the Schwarzschild radius. The tortoise coordinate is then defined as
\begin{align}
z&=- \int^r \frac{dr'}{f(r')}\sim -\frac{\beta_0}{4\pi}\log\left(1-\frac{r_s}{r}\right),\hspace{10 mm}r\to r_s.\label{tortoisebh}
\end{align}
Comparing (\ref{tortoisechain}) with (\ref{tortoisebh}) leads to the identification
\begin{align}
n\sim \frac{1}{\sqrt{1-\frac{r_s}{r}}},\hspace{10 mm}r\to r_s.
\end{align}
Accordingly, the large $n$ limit corresponds to the near horizon region.\\
\indent The equation of motion (\ref{eomcont}) is free, but we will see later that it is possible to define a scattering potential $V(n)$. Assuming that the $b_n$'s have an expansion in integer powers of $1/n$, the potential can be expanded at large $n$ as
\begin{align}\label{vinteger}
V(n)=\sum_{k=1}^{\infty}V_k n^{-k}\sim \sum_{k=1}^{\infty}V_k \left(1-\frac{r_s}{r}\right)^{\frac{k}{2}}.
\end{align}
This matches the form of the potential in the Teukolsky equation \cite{Teukolsky:1973ha}, where terms with odd integers $k$ in the sum appear for fermions \cite{Khriplovich:2005wf}. The power series (\ref{vinteger}) is the simplest possibility, but we will see that more exotic potentials can appear as well. Heuristically, these correspond to stringy or quantum corrections to the smooth black hole potential. Related ideas appear in \cite{Gesteau:2023rrx,Gesteau:2024rpt,Ouseph:2023juq,Grozdanov:2016vgg,Grozdanov:2018gfx,Nebabu:2023iox}, and other connections between black holes and operator growth are discussed in \cite{Susskind:2018pmk,Susskind:2018tei,Susskind:2019ddc,Susskind:2020gnl,Brown:2018kvn,Ambrosini:2024sre,Rabinovici:2023yex,Lin:2022rbf,Brown:2018bms,Lin:2019qwu}.
\section{Green's functions and quasinormal modes}\label{greenssec}
\indent In this section we turn to the real time dynamics, and explain how to compute the retarded Green function,
\begin{align}
G_R(t)=-iG_+(t)\theta(t).
\end{align}
Our starting point is the continued fraction representation of $G_R$ in the upper half plane \cite{Viswanath1994TheRM,Nandy:2024htc},
\begin{align}\label{uhp}
G_R(\omega)=\left(\mathcal{O}\left|\frac{1}{\omega-\mathcal{L}}\right|\mathcal{O}\right)=\frac{1}{\omega-\frac{b_1^2}{\omega-\frac{b_2^2}{\omega-\ldots}}},\hspace{10 mm}\text{Im }\omega>0.
\end{align}
The numerator and denominator of $G_R$ can be computed as follows. Let $g_\omega$ and $\tilde{g}_\omega$ be linearly independent solutions to (\ref{eom}) satisfying the initial conditions
\begin{align}\label{gboundary}
g_\omega(-1)=\tilde{g}_\omega(0)=0,\hspace{5mm }\tilde{g}_{\omega}(-1)=g_{\omega}(0)=1.
\end{align}
We also define $b_0=i$. Then it is straightforward to check that
\begin{align}\label{naive}
G_R(\omega)=\lim_{n\to\infty}\frac{\tilde{g}_\omega(n)}{g_\omega(n)}.
\end{align}
\indent Although the continued fraction (\ref{uhp}) converges in the upper half plane, it is not useful for $\text{Im }\omega\le 0$. In order to write down a more general expression, we draw inspiration from holography \cite{Maldacena:1997re}, where one computes boundary correlators as extrapolated bulk correlators \cite{Son:2002sd}. In the present context, we identify the bulk radial direction with $n$, and the boundary is located at $n=0$. Our task is therefore to construct a bulk retarded Green's function $G_R(\omega,n,n')$, which satisfies the difference equation
\begin{align}
i\omega G_R(\omega,n,n')-b_{n+1}G_R(\omega,n+1,n')+b_{n}G_R(\omega,n-1,n')=i(-1)^{n}\delta_{n,n'}.
\end{align}
The solution can be obtained by a straightforward generalization of scattering theory to the discrete setting, and we closely follow the presentation of \cite{newton_scattering_theory_of_waves_and_particles,Festuccia:2005pi,Festuccia:2008zx,festucciathesis} in what follows.\\
\indent In order to understand the proper boundary conditions for the retarded Green's function at $n=\infty$, we take the large $n$ limit of the equations of motion,
\begin{align}
i\omega\phi_\omega(n)=\frac{\pi}{\beta_0} \left((n+1)\phi_\omega(n+1)-n\phi_\omega(n-1)\right).
\end{align}
The two linearly independent solutions are
\begin{align}\label{hrha}
h^R_\omega(n)\sim n^{-\frac{1}{2}+\frac{i\omega\beta_0}{2\pi}},\hspace{10 mm}h^A_\omega(n)\sim(-1)^nn^{-\frac{1}{2}-\frac{i\omega\beta_0}{2\pi}},\hspace{10 mm}n\to \infty.
\end{align}
The retarded solution $h^R_\omega(n)$ is ingoing at the horizon, and the bulk retarded Green's function should satisfy these ingoing asymptotics. In addition, $G_R(\omega,n,n')$ should vanish at $n=-1$. The appropriate solution is therefore
\begin{align}
G_R(\omega,n,n')=
\begin{cases}
-\frac{ig_\omega(n) h^R_\omega(n')}{f(\omega)},\hspace{5 mm}n'\ge n\\
-\frac{ih^R_\omega(n)g_\omega(n')}{f(\omega)} ,\hspace{5 mm}n\ge n'.
\end{cases}
\end{align}
Here we have defined the Jöst function 
\begin{align}\label{jostdef}
f(\omega)=W( h_\omega^R,g_\omega),
\end{align}
where the Wronskian is given by
\begin{align}
W(\phi_1,\phi_2)=(-1)^n b_{n+1}(\phi_1(n+1)\phi_2(n)-\phi_2(n+1)\phi_1(n)).
\end{align}
It is straightforward to check that the Wronskian of any two solutions is conserved, so that $f(\omega)$ is independent of $n$. \\
\indent The boundary correlator is obtained by extrapolation of $G_R(\omega,n,n')$ to $n=n'=0$. We find
\begin{align}\label{correct}
G_R(\omega)=-\frac{K(\omega)}{f(\omega)}=-\frac{h^R_\omega(0)}{h^R_\omega(-1)},
\end{align}
where 
\begin{align}
K(\omega)=-W(h_\omega^R,\tilde{g}_\omega).
\end{align}
When evaluating the denominator on the right hand side of (\ref{correct}), one should recall that $b_0=i$. \\
\indent We can now check that (\ref{correct}) reduces to (\ref{naive}) in the upper half plane. To do so, note the relations
\begin{align}
g_\omega(n)&=\frac{\beta_0}{2\pi}\left(f(-\omega)h^R_\omega(n)+f(\omega)h^A_\omega(n)\right)\label{gcomb}\\
\tilde{g}_\omega(n)&=\frac{\beta_0}{2\pi}\left(K(-\omega)h^R_\omega(n)-K(\omega)h^A_\omega(n)\right).
\end{align}
For $\text{Im }\omega>0$, the advanced solution is dominant, and 
\begin{align}
\lim_{n\to \infty}\frac{\tilde{g}_\omega(n)}{g_\omega(n)}=-\frac{K(\omega)}{f(\omega)}.
\end{align}
This matches the result (\ref{correct}). But (\ref{correct}) is more general than (\ref{naive}), and we will see in Section \ref{spectralsec} that (\ref{correct}) provides an analytic continuation of the correlator into the lower half plane when the Lanczos coefficients are sufficiently smooth.
\\\indent At certain special frequencies, it might be the case that the retarded solution vanishes at $n=-1$. This boundary condition defines a quasinormal mode, in direct analogy with holography \cite{Horowitz:1999jd,Berti:2009kk}.  It follows from (\ref{correct}) that quasinormal mode frequencies are singularities of the retarded Green's function, which are also known as Pollicott-Ruelle resonances \cite{Pollicott1985,PhysRevLett.56.405}. These resonances were studied in the quantum context in \cite{Prosen_2002,PROSEN2004244,dodelson2024ringdownsykmodel,zhang2024thermalization,Mori:2023qbd,Teretenkov:2024uwm,Banuls:2019qrq}. Since $G_R(\omega)$ is analytic in the upper half plane, the quasinormal modes must satisfy $\text{Im }\omega<0$.\\
\indent It will be useful later to express the spectral density in terms of the Jöst function. We note that 
\begin{align}
h^R_\omega(n)&=-i(K(\omega)g_\omega(n)+f(\omega)\tilde{g}_{\omega}(n))
\label{hrcomb}.
\end{align}
Plugging (\ref{hrcomb}) into (\ref{gcomb}), we find the relation 
\begin{align}
K(\omega)f(-\omega)+K(-\omega)f(\omega)=\frac{2\pi i }{\beta_0}.
\end{align}
Here we used the properties 
\begin{align}
h^A_{\omega}(n)&=(-1)^nh^R_{-\omega}(n)\\
g_\omega(n)&=(-1)^n g_{-\omega}(n)\\
\tilde{g}_\omega(n)&=(-1)^{n+1} \tilde{g}_{-\omega}(n).
\end{align}
Using $(K(\omega))^*=-K(-\omega)$ and $(f(\omega))^*=f(-\omega)$, the spectral density becomes
\begin{align}\label{spectraljost}
G_+(\omega)=-2\text{Im }G_R(\omega)=\frac{2\pi}{\beta_0f(\omega)f(-\omega)}.
\end{align}
\section{The free theory}\label{freesec}
Let us now work out an example which arises in two dimensional conformal field theory \cite{Dymarsky:2021bjq} and in the large $q$ limit of the SYK model \cite{Parker:2018yvk,Tarnopolsky:2018env}. Since this example can be solved explicitly, we will use it as the starting point for a perturbative expansion. For this reason, we refer to it as the free theory.\\
\indent The two point function reads 
\begin{align}\label{twopointlargeq}
G_+(t)=(\sech t)^\eta.
\end{align}
Here we have set $\beta_0=\pi$, but it can be easily restored using dimensional analysis. In the case of 2D CFT, one should use the Wightman inner product \cite{Parker:2018yvk} instead of (\ref{innerprod}) to recover the two point function (\ref{twopointlargeq}). The Lanczos coefficients are smooth functions of $n$ which can be computed exactly,
\begin{align}\label{freebns}
b_n=\sqrt{n(n-1+\eta)}.
\end{align}
Given these Lanczos coefficients, we would now like to reproduce the Green's functions from the black hole scattering problem.\\
\indent First we construct the solutions $g_\omega(n)$ and $h^R_\omega(n)$, which are related to the Meixner polynomials \cite{Parker:2018yvk}. The answer is \cite{Muck:2022xfc}
\begin{align}
g_\omega(n)&=\sqrt{\frac{\Gamma(n+\eta)}{\Gamma(n+1)\Gamma(\eta)}}{_2F_1}\left(-n,\frac{\eta-i\omega}{2},\eta,2\right)\\
h_\omega^R(n)&=\frac{\sqrt{\Gamma(n+\eta)\Gamma(n+1)}}{\Gamma\left(n+1+\frac{\eta-i\omega}{2}\right)}{_2F_1}\left(1-\frac{\eta+i\omega}{2},\frac{\eta-i\omega}{2},n+1+\frac{\eta-i\omega}{2},\frac{1}{2}\right).\label{hr0}
\end{align}
Computing the Wronskian (\ref{jostdef}), we find the Jöst function
\begin{align}
f(\omega)=\frac{2^{1-\frac{\eta+i\omega}{2}}\sqrt{\Gamma(\eta)}}{\Gamma\left(\frac{\eta-i\omega}{2}\right)}.
\end{align}
The relation (\ref{spectraljost}) then gives the spectral density
\begin{align}
G_+(\omega)=\frac{\Gamma\left(\frac{\eta-i\omega}{2}\right)\Gamma\left(\frac{\eta+i\omega}{2}\right)}{2^{1-\eta}\Gamma(\eta)},
\end{align}
which matches the Fourier transform of (\ref{twopointlargeq}). Notice that $G_+(\omega)$ is meromorphic in frequency, and has no zeroes in the complex plane \cite{Festuccia:2005pi,Festuccia:2008zx,festucciathesis,Dodelson:2023vrw,Grozdanov:2024wgo}. These properties will play an important role in the next section.\\
\indent Note that the quasinormal mode frequencies are 
\begin{align}
\omega_m=-i(\eta+2m),\hspace{10 mm}m=0,1,\ldots
\end{align}
At these values of $\omega$, the hypergeometric functions become polynomials in $n$, and the wavefunctions $g_\omega(n)$ simplify accordingly.
\section{Analytic structure of the spectral density}\label{spectralsec}
\indent We are now ready to study the analytic properties of the Green's functions in the complex frequency plane. The strategy is to develop a convergent perturbative expansion around the free theory discussed in the previous section.
\subsection{A Dyson series for the retarded solution}
 It follows from (\ref{spectraljost}) that the analytic properties of the spectral density are determined by those of the Jöst function. In fact, one can make a stronger statement. The solution $g_\omega$ is automatically analytic in frequency, since it is a polynomial in $\omega$. Since $f(\omega)$ is given by the Wronskian of $h^R_\omega$ with $g_\omega$, it therefore suffices to study the analytic structure of the retarded solution.\\
\indent Let us refer to the coefficients and retarded solution of the free theory as $b^0_n$ and $h^{R0}_\omega$. It is useful to make the change of variables \cite{GERONIMO1988251}
\begin{align}
\hat{\phi}_\omega(n)=\alpha_n\phi_\omega(n),\hspace{10 mm}\alpha_n=\prod_{i=1}^{n}\frac{b_n}{b^0_n},
\end{align}
after which the equations of motion (\ref{eom}) become 
\begin{align}
b^0_{n+1}\hat{\phi}_\omega(n+1)-\frac{b_n^2}{b^0_n}\hat{\phi}_\omega(n-1)=i\omega\hat{\phi}_{\omega}(n).
\end{align}
Then we can write a discrete integral equation for the retarded solution $\hat{h}^R_\omega$ with coefficients $b_n$ \cite{GERONIMO1988251},
\begin{align}\label{dysonseries}
\hat{h}^{R}_{\omega}(n)=h^{R0}_{\omega}(n)-i\sum_{n'=n}^{\infty}(-1)^{n'}G^0_\omega (n,n'+1)V(n')\hat{h}^{R}_{\omega}(n').
\end{align}
Here the potential takes the form
\begin{align}\label{potential}
V(n)=b^0_{n+1}\left(\left(\frac{b_{n+1}}{b^0_{n+1}}\right)^2-1\right),
\end{align}
and the unperturbed Green's function satisfies 
\begin{align}
i\omega G^0_\omega(n,n')-b^0_{n+1}G^0_\omega(n+1,n')+b^0_nG^0_\omega(n-1,n')=i(-1)^n\delta_{n,n'},
\end{align}
with the boundary condition 
\begin{align}
G^0_\omega(n,n')&=0,\hspace{10 mm}n\ge n'.
\end{align}
Setting $\beta_0=\pi$ as above, the solution is 
\begin{align}
G_\omega^0(n,n')=\frac{i}{2}(h^{R0}_\omega(n)h^{A0}_\omega(n')-h^{A0}_\omega(n)h_\omega^{R0}(n')),\hspace{10 mm}n\le n'.
\end{align}
\indent The equation (\ref{dysonseries}) can be solved by iteration. For instance, to first order we have 
\begin{align}\label{firstorder}
\hat{h}^{R}_{\omega}(n)=h^{R0}_{\omega}(n)-i\sum_{n'=n}^{\infty}(-1)^{n'}G^0_\omega (n,n'+1)V(n')h^{R0}_{\omega}(n')+\ldots
\end{align}
In order to check the convergence of these iterations, note the following bounds
\begin{align}\label{bounds}
|h_\omega^{R0}(n)|&<C(\omega)n^{-\frac{1}{2}-\frac{\text{Im }\omega}{2}}\\
|G^0_\omega(n,n'+1)|&<C(\omega)\left(\frac{n'}{n}\right)^{-\frac{1}{2}+\frac{|\text{Im }\omega|}{2}},\label{bounds2}
\end{align}
for $n,n'>0$ and some constant $C(\omega)>0$. It is clear that these bounds hold asymptotically, and away from large $n$ and $n'$ they can be checked numerically. Plugging the bounds (\ref{bounds} and (\ref{bounds2}) into the integral equation (\ref{dysonseries}) and using the Cauchy-Schwarz inequality as in \cite{newton_scattering_theory_of_waves_and_particles,Hartnoll:2005ju}, we find
\begin{align}\label{csinequality}
|\hat{h}^R_\omega(n)|<C(\omega)n^{-\frac{1}{2}-\frac{\text{Im }\omega}{2}}\exp\left(C(\omega)\sum_{n'=n}^{\infty}|V(n')|\left(\frac{n'}{n}\right)^{-1+\frac{|\text{Im }\omega|-\text{Im }\omega}{2}}\right).
\end{align}
\indent It follows from (\ref{csinequality}) that the series converges as long as 
\begin{align}
\sum_{n'=1}^{\infty}|V(n')|n'^{-1+\frac{|\text{Im }\omega|-\text{Im }\omega}{2}}<\infty.
\end{align}
In terms of the Lanczos coefficients, this  becomes
\begin{align}\label{condition}
\sum_{n'=1}^{\infty}|b_{n'}-b^0_{n'}|n'^{-1+\frac{|\text{Im }\omega|-\text{Im }\omega}{2}}<\infty.
\end{align}
Next we will consider the implications of this condition in several examples.
\subsection{Examples of Lanczos asymptotics}
\indent First suppose that the Lanczos coefficients have a convergent expansion in powers of $1/n$, 
\begin{align}\label{asympsmooth}
b_n=n+\frac{\eta-1}{2}+\frac{c}{n^\gamma}+\ldots 
\end{align}
We assume that $\eta>0$, so that the free coefficients $b^0_n$ are well-defined. Then we can perturb around the free theory with parameter $\eta$, and the sum (\ref{condition}) converges in the upper half plane provided that $\gamma>0$. In addition, the perturbative expansion converges in a strip of the lower half plane,
\begin{align}\label{strip}
\text{Im }\omega>-\gamma.
\end{align}
\indent The condition (\ref{strip}) should not be understood to mean that the solution necessarily blows up for $\text{Im }\omega<-\gamma$, and in certain cases the solutions can be extended further \cite{newton_scattering_theory_of_waves_and_particles}. Indeed, plugging (\ref{asympsmooth}) into the first order answer (\ref{firstorder}) and taking the large $n,n'$ limit of the summand, we find
\begin{align}
\hat{h}_\omega^R(n)-h_\omega^{R0}(n)&\propto \sum_{n'=n}^{\infty}\left((-1)^{n'}n'^{-1-\gamma+i\omega}h^{A0}_{\omega}(n)+n'^{-1-\gamma}h_\omega^{R0}(n)\right)\\
&=\frac{n^{-\frac{1}{2}-\frac{i\omega}{2}}}{2^{1+\gamma-i\omega}}\left(\zeta\left(1+\gamma-i\omega,\frac{n}{2}\right)-(n\to n+1)\right)+\zeta(1+\gamma,n)n^{-\frac{1}{2}+\frac{i\omega}{2}}\notag.
\end{align}
This is an entire function of $\omega$. The same conclusion seems to hold at higher orders in perturbation theory, although we do not have a proof. Assuming that there is no divergence arising from the sum over iterations, it follows that the spectral density is a meromorphic function of frequency with no zeroes in the complex plane. Holographic models provide one class of theories that satisfy this constraint, but there are also examples that are not maximally chaotic such as the large $q$ SYK chain \cite{Choi:2020tdj,Dodelson:2023vrw}.\\
\indent The next situation we will consider is when alternating terms are allowed in the asymptotic expansion of $b_n$, so that the potential contains a term of the form
\begin{align}
V(n)\sim \frac{(-1)^n}{n^\gamma}.
\end{align}
Repeating the above calculation, one finds a simple pole in $\hat{h}^R_\omega(n)$, 
\begin{align}
\hat{h}^R_\omega(n)\propto \frac{1}{\omega+i\gamma}.
\end{align}
As a consequence, the spectral density is still meromorphic, but has zeroes in the complex frequency plane. An example of such a theory is SYK at finite $q$ \cite{dodelson2024ringdownsykmodel}. Note that $\gamma$ can be taken to be complex, provided that the complex conjugate term is added to make the potential real. Therefore the zeroes come in complex conjugate pairs.\\
\indent Finally, it is not difficult to concoct terms in the potential that lead to branch cuts. The prototype is 
\begin{align}
V(n)\sim \frac{(-1)^n}{n^\gamma(\log n)^{\nu}},
\end{align}
which yields a branch cut of the form 
\begin{align}
\hat{h}^R_\omega(n)\propto \frac{1}{(\omega+i\gamma)^{1-\nu}}.
\end{align}
Similar comments were made in \cite{Bhattacharjee:2022vlt}. In the next section we will provide evidence that branch cuts arise in the mass-deformed SYK theory.
\section{Two case studies in SYK}\label{examplesec}
In this section we present two instructive examples within the context of the SYK model at infinite temperature. The Hamiltonian is \cite{Maldacena:2016hyu,kitaev,Sachdev_1993,Polchinski:2016xgd} 
\begin{align}
H_q=i^{q/2}\sum_{1\le i_1\le\ldots\le i_q\le N}j_{i_1\ldots i_q}\psi_{i_1}\cdots \psi_{i_q},\hspace{10 mm}\langle j^2_{i_1\ldots i_q}\rangle=\frac{(q-1)!\mathcal{J}_q^2}{2^{1-q}qN^{q-1}}.
\end{align}
From now on we work in units where $\mathcal{J}_q=1$.

\subsection{$q=4$}
\begin{figure}[t]
 \begin{subfigure}{.55\textwidth}
\includegraphics[width=\linewidth,valign=c]{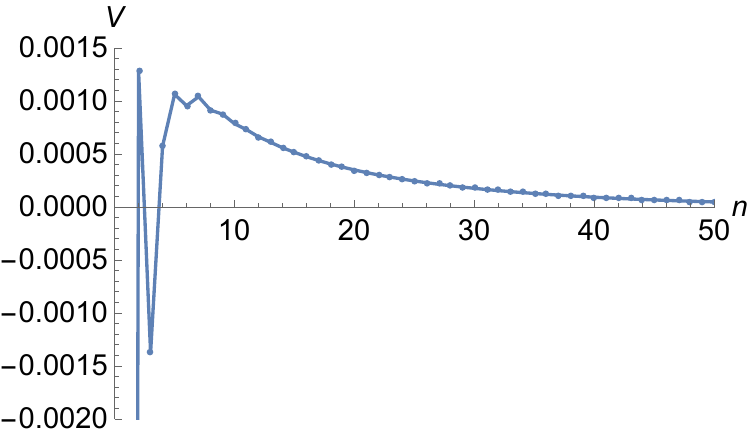}
\caption{\label{qeq4lanc}}
    \end{subfigure}\hfill
 \begin{subfigure}{0.35\textwidth}
    \centering
    \includegraphics[width=\linewidth,valign=c]{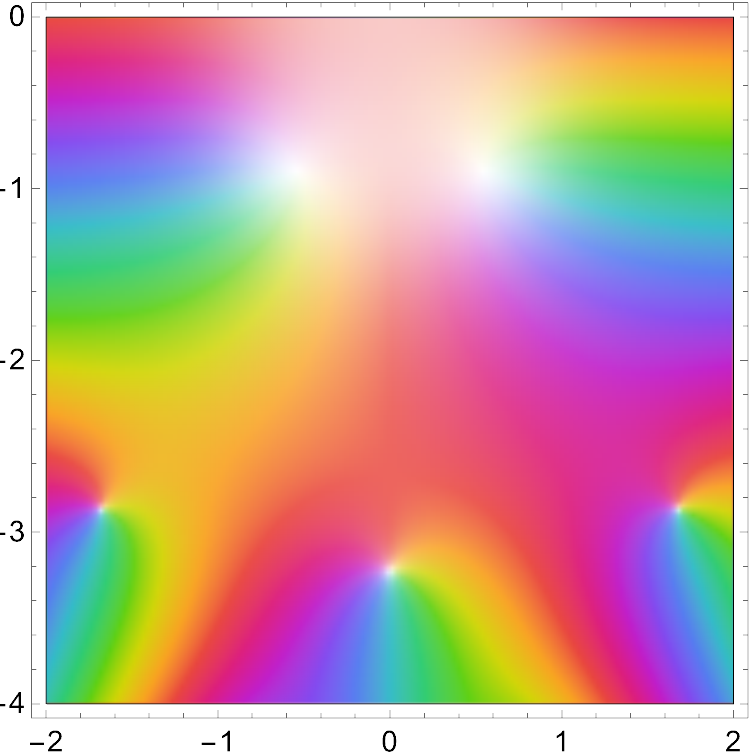}
  \caption{\label{specqeq4}}
  \end{subfigure} 
    \caption{The scattering potential and spectral density for $q=4$ SYK. The colors in the right plot signify the phase, and the poles are marked by the white dots.}
  \end{figure}

For finite $q$ the Lanczos coefficients can be computed exactly using the Schwinger-Dyson equation \cite{Parker:2018yvk,dodelson2024ringdownsykmodel}. Working with the first 100 $b_n$'s, we fitted the results to a straight line to extract the appropriate values of $\eta$ and $\beta_0$ for the free theory (\ref{freebns}), and computed the potential using (\ref{potential}). As shown in Figure \ref{qeq4lanc}, the potential decays as $1/n$ at large $n$, so we expect perturbation theory to converge in a region of the lower half plane.\\
\indent In order to compute the correlator, we fit the Lanczos coefficients to a power series at large $n$, 
\begin{align}\label{largenqeq4}
b_n\sim .623 n+\frac{.0064}{n}+\frac{.001}{n^2}-\frac{.07}{n^3}+\ldots
\end{align}
We then solve the difference equation (\ref{eom}) for the retarded solution perturbatively in $1/n$. Lastly, we compute the solution $g_\omega(n)$ exactly using (\ref{eom}) along with the boundary conditions (\ref{gboundary}), and evaluate its Wronskian with $h^R_\omega(n)$ at large $n$. The spectral density is given in terms of the Wronskian by (\ref{spectraljost}).\\
\indent The resulting spectral density is shown in Figure \ref{specqeq4}. We see that there are five isolated poles in the lower half frequency plane. The location of the pole closest to the real axis matches the answer found in \cite{Roberts:2018mnp,dodelson2024ringdownsykmodel} to six digits, and the other poles have converged to two digits. In order to push further down into the complex plane, one would need to fit higher order terms in the expansion (\ref{largenqeq4}). Since the spectral density has zeroes \cite{dodelson2024ringdownsykmodel}, we expect from the discussion in the previous section that there are terms in the expansion that oscillate as $(-1)^n/n^\gamma$. Such oscillations are indeed visible numerically, but we were not able to fit them to high accuracy.\\
\indent Note from Figure \ref{qeq4lanc} that the potential is numerically small. This means that we expect perturbation theory to work well. Indeed, let us suppose that we have access to only the first four Lanczos coefficients: 
\begin{align}
b_1=\frac{1}{\sqrt{2}},\hspace{5 mm}b_2=\sqrt{\frac{3}{2}},\hspace{5 mm}b_3=\sqrt{\frac{7}{2}},\hspace{5 mm}b_4=\sqrt{\frac{87}{14}}.
\end{align}
Fitting $b_3$ and $b_4$ to a straight line gives $\beta_0\sim 5.05$ and $\eta\sim 1.02$. We then approximate the Wronskian as 
\begin{align}
f(\omega)&\approx b_4 (h^{R0}_\omega(3)g_\omega(4)-h^{R0}_\omega(4)g_\omega(3)),
\end{align}
where $h^{R0}_\omega$ is the free wavefunction (\ref{hr0}). The root of the right hand side with smallest imaginary part is
\begin{align}
\omega_{\pm}\approx\pm .544-.897i,
\end{align}
which matches the correct answer \cite{Roberts:2018mnp,dodelson2024ringdownsykmodel} with less than percent error. This is a striking result: the long time behavior of the correlator can be computed to high accuracy from its first four time derivatives at the origin. This is only possible because the Lanczos coefficients immediately converge to the asymptotic linear behavior in this model, and we do not expect the same level of accuracy in a generic situation. \\
\indent The fast convergence that we have observed is also a feature of the cycle expansion of Ruelle resonances in classical chaotic systems \cite{ChaosBook,Gaspard_1998,PhysRevLett.61.2729}. The basic mechanism underlying the convergence in that setting is the shadowing of long unstable periodic orbits by short orbits. In our context, it is natural to identify simple and complex operators with short and long orbits respectively. It would be interesting to make this precise.
\begin{figure}[t]
 \begin{subfigure}{.55\textwidth}
\includegraphics[width=\linewidth,valign=c]{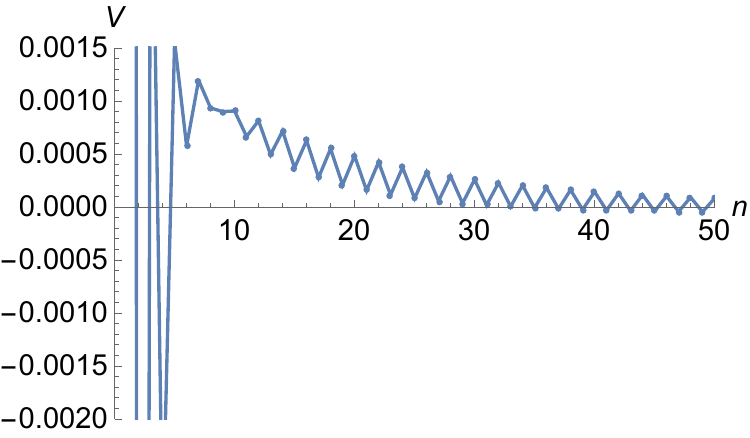}
\caption{\label{potmass}}
    \end{subfigure}\hfill
 \begin{subfigure}{0.35\textwidth}
    \centering
    \includegraphics[width=\linewidth,valign=c]{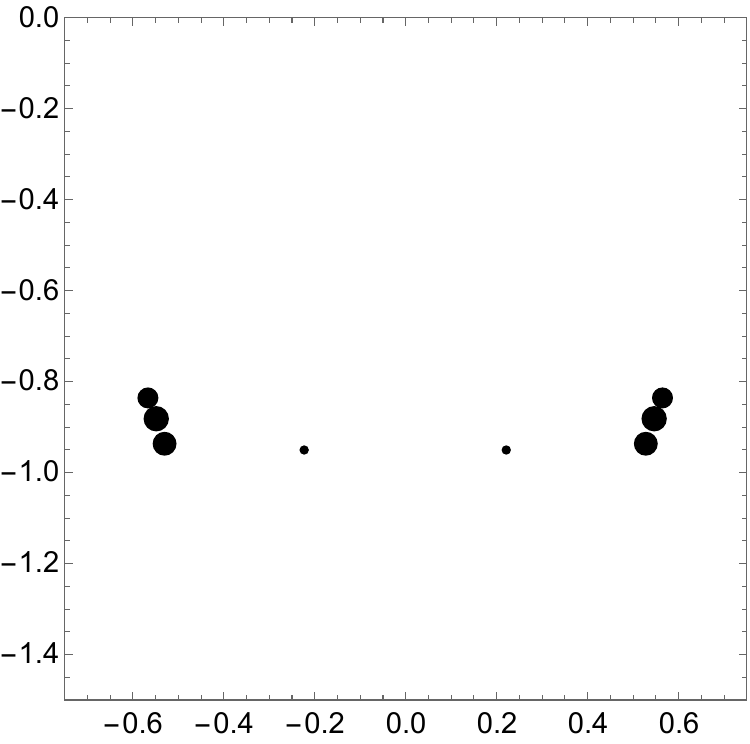}
  \caption{\label{specmass}}
  \end{subfigure} 
    \caption{The potential and spectral density for mass-deformed SYK, with $\lambda=1/150$ and $\lambda=1/1000$ respectively.}
  \end{figure}
\subsection{Mass deformation}
Our second example is the mass-deformed theory \cite{Garcia-Garcia:2017bkg}, 
\begin{align}\label{massham}
H=H_4+\lambda H_2.
\end{align}
At low energies this model is integrable, and at high energies it reduces to the ordinary $q=4$ theory. The spectral density in the $q=2$ model can be computed exactly \cite{Maldacena:2016hyu}, 
\begin{align}\label{qeq2g}
G_+(\omega)=\sqrt{1-\frac{\omega^2}{4}}\theta\left(1-\frac{\omega^2}{4}\right).
\end{align}
This function has compact support, and cannot be analytically continued into the complex plane. As $\lambda$ is dialed up, we expect that the Green's function in the mass-deformed theory should interpolate between the meromorphic $q=4$ result and the nonanalytic answer (\ref{qeq2g}).  \\
\indent  For any nonzero $\lambda$, we find that the Lanczos coefficients are staggered as a function of $n$, which leads to oscillations in the potential. This is depicted in Figure \ref{potmass}. As we found it difficult to reliably fit the $b_n$'s to an asymptotic series, we instead used the approach of \cite{dodelson2024ringdownsykmodel} to compute the spectral density. The locations of the poles are shown in Figure \ref{specmass} for $\lambda=1/1000$, where 800 moments were used in the computation.\\
\indent Evidently, the leading quasinormal mode splits into a series of poles for any nonzero $\lambda$. Our expectation is that the discreteness of these poles is an artifact of truncating the moment expansion, so that pushing to higher orders would lead to an accumulation of poles along a branch cut. We suspect that this cut is a consequence of the integrability of the mass-deformed model at low energies. Indeed, such cuts often appear in classical chaotic systems with integrable regions of phase space \cite{ChaosBook}.

\section{Outlook}
In this work we have formulated the real time dynamics of a chaotic system as a scattering problem on a black hole background. The inputs are a set of static quantities, namely the Lanczos coefficients, and the output is the late time behavior of the thermal two point function. Our hope is that this procedure can be applied to realistic chaotic systems such as the critical Ising model in three dimensions or quantum chromodynamics, where real time thermal quantities are presently inaccessible.\\
\indent The scattering potentials studied in this paper have been relatively featureless, due to the simplicity of the models we have considered. It would be interesting to find a model whose potential has an unstable maximum, which might be interpreted as a photon sphere. We expect that the two point function in such a model should exhibit a bulk cone singularity \cite{Hubeny:2006yu, Dodelson:2020lal,Dodelson:2023nnr}. Similarly, the black hole singularity should be related to divergences of the analytically continued potential in the complex $n$ plane, and one could look for signatures of the black hole singularity in the two point function \cite{Festuccia:2005pi,Fidkowski:2003nf}.\\
\indent There are several basic questions that we have not attempted to address here. First, it is not clear how the scattering potential we have constructed is related to its holographic counterpart when the latter exists. Relatedly, the equation of motion we find is always linear, whereas in holography one expects a nonlinear wave equation in the presence of $1/N$ corrections. Lastly, in this paper we have exclusively worked in the thermodynamic limit. When the entropy is finite, the Lanczos coefficients are expected to eventually plateau \cite{Barbon:2019wsy,Kar:2021nbm}, and our perturbation theory breaks down. We hope to develop techniques to address this regime in future work. 
\section*{Acknowledgments}
We are grateful to Alexander Avdoshkin, Oleg Lychkovskiy, and Pavel Nosov for helpful discussions, and to Anatoly Dymarsky and Robin Karlsson for comments on the draft. This work was supported by DOE grant DE-SC/0007870 and the Frankel-Goldfield Research Fund.

\appendix

\bibliographystyle{JHEP}
\bibliography{mybib}
  
\end{document}